\documentclass[twocolumn,amsmath,amssymb,nofootinbib,amsfonts,showkeys]{revtex4}
\usepackage{graphicx}
\usepackage{bm}
\usepackage{mathtext}
\usepackage[english]{babel}

\begin{document}

\title{Dynamics of expansion of the Universe \\
in model with the additional coupling \\
between dark energy and dark matter}
\author{R. Neomenko\footnote{oz.rik@hotmail.com}, B. Novosyadlyj\footnote{novos@astro.franko.lviv.ua}, O. Sergijenko\footnote{olka@astro.franko.lviv.ua}}
\affiliation{Astronomical Observatory of Ivan Franko National University of Lviv, Kyryla i Methodia str., 8, Lviv, 79005, Ukraine}

\begin{abstract}

We study the dynamics of expansion of the homogeneous isotropic Universe and the evolution of its components in the model with nonminimally coupled dynamical dark energy. Dark energy, like the other components of the Universe, is described by the perfect fluid approximation with the equation of state (EoS) $p_ {de}=w\rho_{de}$, where the EoS parameter $w$ depends on time and is parameterized via the squared adiabatic sound speed $c_{ a}^2$ which is assumed to be constant. On basis of the general covariant conservation equations for the interacting dark energy and dark matter and Einstein equations in Friedmann-Lemaitre-Robertson-Walker metric we analyze the evolution of energy densities of the hidden components and the dynamics of expansion of the Universe with two types of interaction: proportional to the sum of densities of the hidden components and proportional to their product. For the first interaction the analytical expressions for the densities of dark energy and dark matter were obtained and analyzed in detail. For the second one the evolution of densities of hidden components of the Universe was analyzed on basis of the numerical solutions of their energy-momentum conservation equations. For certain values of the parameters of these models the energy densities of dark components become negative. So to ensure that the densities are always positive we put constraints on the interaction parameter for both models.
\end{abstract}

\keywords{nonminimally coupled dark energy, dynamics of expansion of the Universe}

\maketitle

\section*{Introduction}

The discovery of accelerated expansion of the Universe has led to the emergence of various hypotheses that explain this phenomenon. It is believed that the acceleration is caused by some unknown component of the Universe, which has a positive energy density and negative pressure. This component is called dark energy. There are various models of dark energy, the simplest of which is the model with the constant energy density and pressure, which is described by the cosmological constant $\Lambda$ in Einstein equations. More complex are the models of dynamical dark energy, in which the parameters depend on time. This type includes different models of a scalar field. To describe the dark energy as well as other components of the Universe the model of perfect fluid is often used. Some models assume the additional interaction of dark energy with other components, which is beyond known 4 physical interactions -- electromagnetic, strong, weak and gravitational
\cite{Amendola2000,Amendola2007,Amendola2010,Bolotin2013,Caldera2009,Gumjudpai2005,Pourtsidou2013,Guo2007,Zimdahl2001}. Since such impact of dark energy on baryon matter or radiation
is not experimentally registered, it is obvious that it is either too weak or absent at all. There are currently no direct experimental constraints on its action on dark matter, since both these components are detected by their gravitational influence on the baryonic matter at cosmological scales. Therefore, in this paper we analyze the effect of the additional interaction of dark energy and dark matter on the dynamics of expansion of the Universe and deduce the constraits on the parameter of such interaction on basis of the null energy condition.

In the considered model of dynamical dark energy the pressure and energy density of dark energy are related via the time-dependent parameter of the equation of state (EoS) $w$ as follows: $p_{de}=w\rho_{de}$. Different works propose different dependences of the EoS parameter on time. We parametrize the evolution of $w$ by the square of the adiabatic sound speed $c_{a}^{2}\equiv\dot{p}_{de}/\dot{\rho}_{de}$, where the dot denotes the derivative on time. Such model for the minimally coupled (non-interacting) dark energy was proposed and studied in \cite{Novosyadlyj2010, Novosyadlyj2011, Novosyadlyj2012, Novosyadlyj2013, Sergijenko2015}, observational constraints on its parameters were obtained in \cite{Sergijenko2011, Novosyadlyj2014}.

In this paper we consider the evolution of dark energy and dark matter, which are coupled with each other by the gravitational and non-gravitational (fifth) interactions, and their influence on the dynamics of expansion of the homogeneous isotropic Universe. The mathematical basis of this work is the conservation equations and Einstein equations in the Universe with Friedmann-Lemaitre-Robertson-Walker metric. We consider 2 types of the interaction depending on the energy densities of both dark components: proportional to the sum of dark energy and dark matter densities and proportional to the product of their densities.

\newpage

\section{Model of the nonminimally coupled dark energy}

\begin{figure*}
\centering
\includegraphics[width=0.41\textwidth]{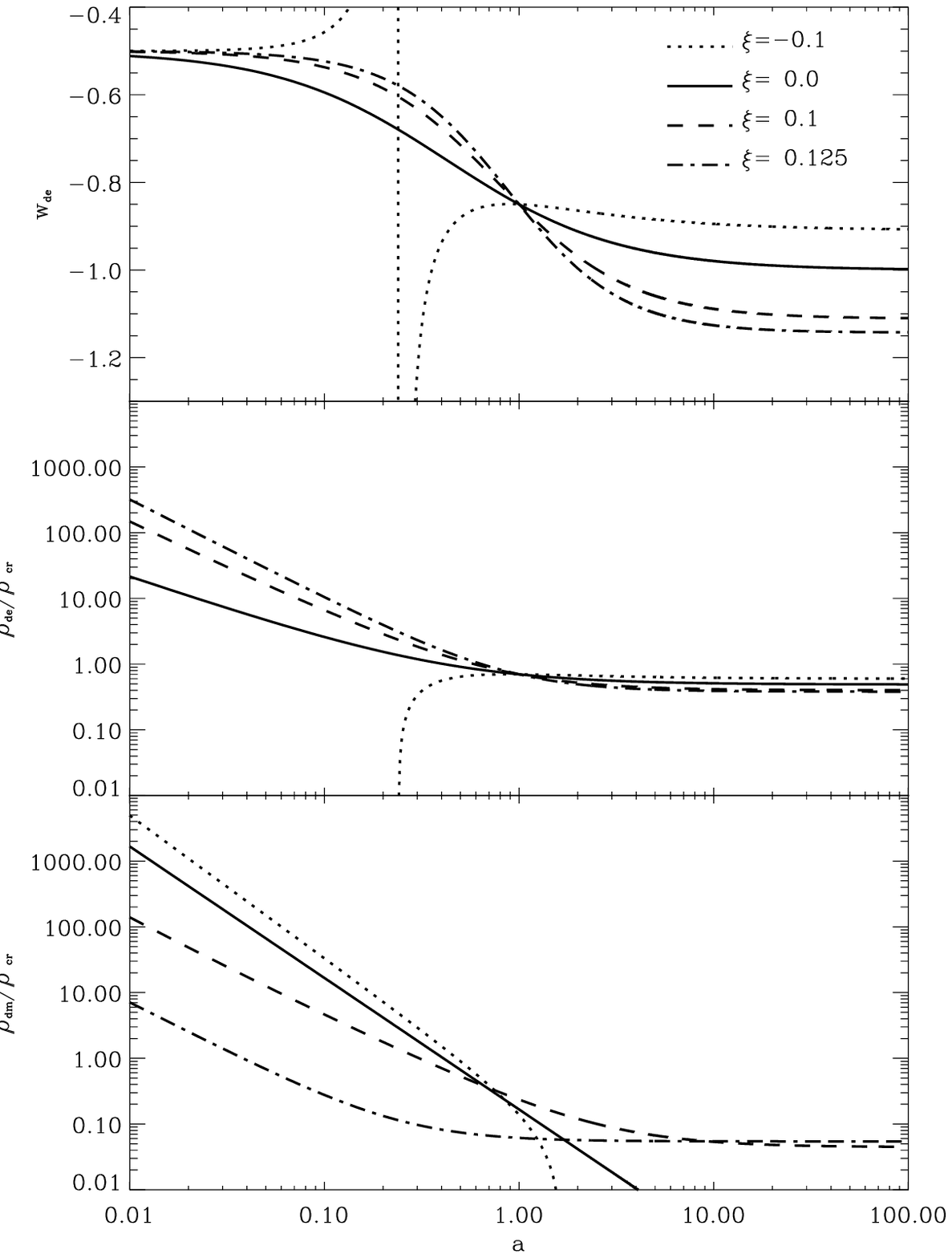}
\includegraphics[width=0.41\textwidth]{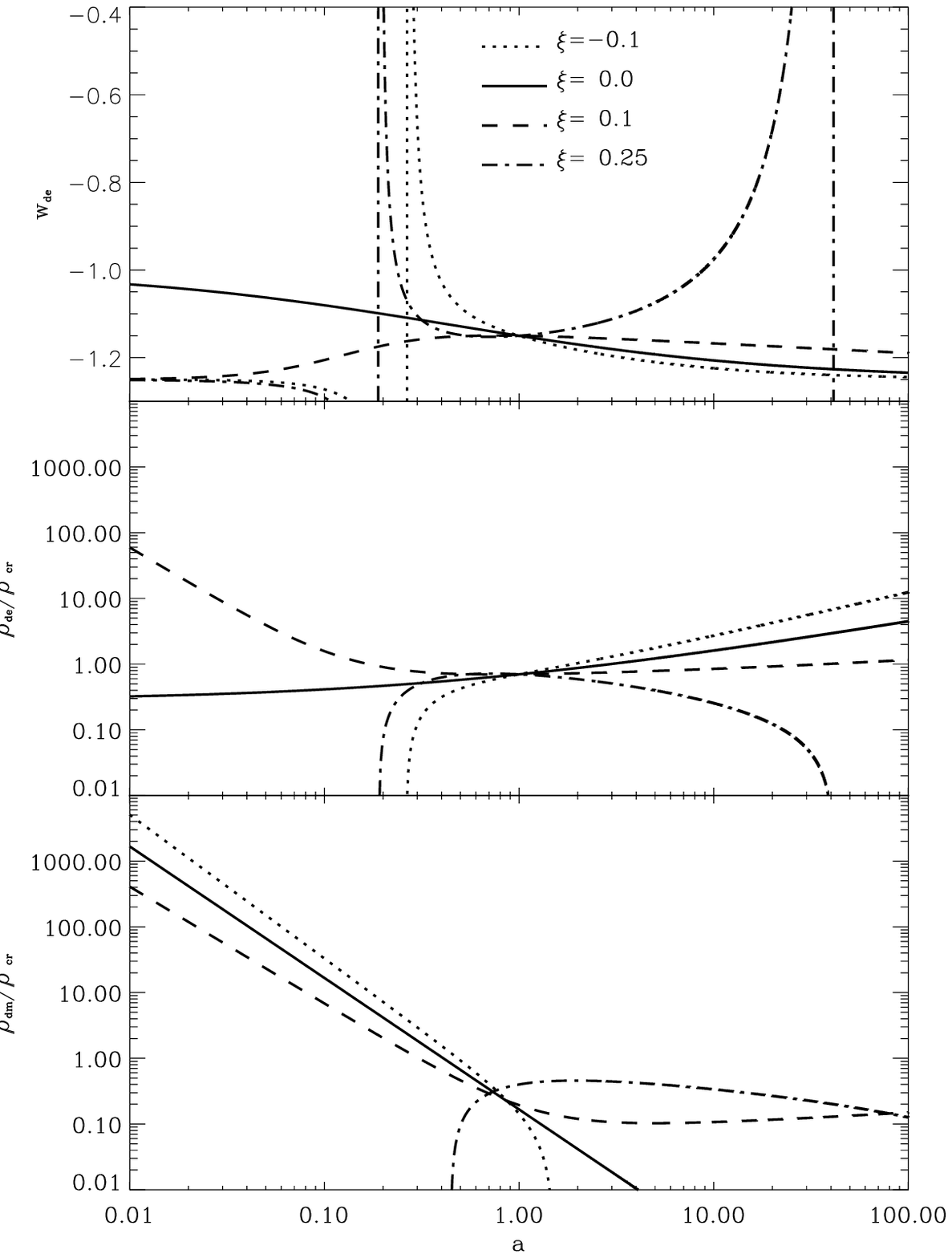}
\caption{Influence of the coupling parameter $\xi$ on dependences of the EoS parameter $w$ and the densities of dark energy and dark matter $\rho_{de}$, $\rho_{dm}$ on $a$. Left --quintessence ($w_{0}=-0.85$, $c_{a}^{2}=-0.5$), right -- phantom ($w_{0}=-1.15$, $c_{a}^{2}=-1.25$). Here $\Omega_{de}=0.7$, $\Omega_{dm}=0.25$, $\Omega_r=4.17\cdot10^{-5}/h^2$, $h\equiv H_0/100$ km$\cdot$с$^{-1}\cdot$Mpc$^{-1}$=0.7.}
  \label{fig:xm sum1}
\end{figure*}

We consider the homogeneous isotropic Universe with the Friedmann-Lemaitre-Robertson-Walker spacetime metric, which in conformal form in spherical coordinates has the form:
\begin{eqnarray}
  & & ds^{2}=g_{ik}dx^idx^k= \nonumber \\
  & & =a^{2}(\eta)[d\eta^{2}-dr^{2}-r^{2}(d\theta^{2}+\sin^{2}\theta d\varphi^{2})], \label{ds}
\end{eqnarray}
where $g_{ik}$ is the metric tensor, $a(\eta)$ is the scale factor describing expansion of the Universe. The conformal time $\eta$ is related to the physical cosmological time $t$ as $cdt=a(\eta)d\eta$, where $c$ is the speed of light. The 3-space of the Universe with the metric (\ref{ds}) is euclidean, that is, a space with zero curvature. We assume that in the modern epoch $a(\eta_0)=1$. We describe each component of the Universe (dark energy ($de$), dark matter ($dm$), baryonic ($b$) and relativistic matter ($r$) -- relic radiation and relic neutrinos) by the perfect fluid approximation in which the energy-momentum tensor is as follows:
\begin{equation}\label{Tik}
T_{i(N)}^{k}=(c^2\rho_{(N)}+p_{(N)})u_{i(N)}u_{(N)}^{k}-p_{(N)}\delta_{i}^{k},
\end{equation}
where $\rho_{(N)}$ is the density of $N$ component, $p_{(N)}$ is its pressure, $u_{i(N)}$ is the 4-vector of velocity. The equation of state for each of the components is as follows: $p_{(N)}=w_{(N)}\rho_{(N)}$. Then from Einstein equations and conservation laws for the interacting dark energy and dark matter we obtain such equations for the dynamics of expansion of the Universe:
\begin{eqnarray}
  & & H^{2}=\frac{8\pi G}{3}\sum_N\rho_{(N)}, \label{H2} \\
  & & qH^2=\frac{4\pi G}{3}\sum_N(\rho_{(N)}+3p_{(N)}), \label{qH2}
\end{eqnarray}
where $H\equiv({1}/{a}){da}/{dt}={\dot{a}}/{a^2}$ is the Hubble parameter, $q\equiv-[{1}/({aH^2})]{d^2a}/{dt^2}=-{\ddot{a}}/({a^3H^2})+1$ -- the deceleration parameter (hereafter $\left(\,\dot{ }\,\right)\equiv{d}/{d\eta}$). The differential energy-momentum conservation law $T^k_{0; k}=0$ for dark components gives the equations describing the evolution of their densities:
\begin{eqnarray}
  & & \dot{\rho}_{de}+3\frac{\dot{a}}{a}\rho_{de}(1+w)=J_{0}, \label{eq de} \\
  & & \dot{\rho}_{dm}+3\frac{\dot{a}}{a}\rho_{dm}=-J_{0}, \label{eq dm}
\end{eqnarray}
where the function $J_{0}$ describes the coupling of dark components and has the dimension of energy density per unit of time. Here and below $w_{de}\equiv w$ and $w_{dm}=0$.

For the baryonic and relativistic components in the post-recombination epoch these equations with $J_0=0$ and $w_b=0$ or $w_r=1/3$ give the well-known dependences for energy densities of these components: $\rho_b(a)=\rho_b^{(0)}a^{-3}$ and $\rho_r(a)=\rho_r^{(0)}a^{-4}$, here and below $(0)$ denotes the current value ($a(\eta_0)=1$).

For models with the nonminimally coupled dark energy it is usually assumed that $w = const$. In this paper, as in the previous one \cite{Neomenko2016}, we consider the nonminimally coupled dynamical dark energy with the generalized linear barotropic equation of state for which $c_{a}^{2}\equiv\dot{p}_{de}/\dot{\rho}_{de}=const$ {\cite{Holman2004, Babichev2005, Novosyadlyj2010}}. From this condition and (\ref{eq de}) we obtain the equation for $w(a)$:
\begin{equation}\label{dw}
  \frac{dw}{da}=\frac{3}{a}(1+w)(w-c_{a}^{2})-\frac{J_{0}}{\rho_{de}a^{2}H}(w-c_{a}^{2}).
\end{equation}
To describe the dynamics of expansion of the Universe and the evolution of densities of its components it is necessary to solve the system of equations (\ref{H2}), (\ref{eq de}), (\ref{eq dm}), (\ref{dw}). The relation between $\rho_{de}$ and $w$ follows from the definition $c_{a}^{2}\equiv\dot{p}_{de}/\dot{\rho}_{de}=const$:
\begin{equation}\label{rw}
  w=c_{a}^{2}+\frac{\rho_{de}^{(0)}(w_{0}-c_{a}^{2})}{\rho_{de}}.
\end{equation}
In this paper we consider the interaction between dark energy and dark matter (hereafter DE-DM interaction) proportional to the Hubble parameter $H$ and depending only on the energy densities of both dark components $J_{0}=aHf(\rho_{de}, \rho_{dm})$. We consider 2 types of such interaction: proportional to the sum of dark components densities and proportional to their product:
\begin{eqnarray}
 &&J_{0}=-3\xi aH(\rho_{de}(a)+\rho_{dm}(a)), \label{Jsum}
\end{eqnarray}
\begin{eqnarray}
 &&J_{0}=-3\epsilon aH\frac{\rho_{de}(a)\rho_{dm}(a)}{\tilde{\rho}}, \label{Jpro}
\end{eqnarray}
where $\tilde{\rho}$ has the dimension of energy density and may be constant or time dependent and $\xi$ and $\epsilon$ are dimensionless coupling parameters assumed to be constant. We study the interaction (\ref{Jsum}) due to the simplicity of obtaining analytical expressions for the energy densities of dark components, and the nonlinear interaction (\ref{Jpro}) is interesting because such its type is quite common in nature and is more realistic than linear interactions of type (\ref{Jsum}) or investigated in \cite{Neomenko2016}.

\section{DE-DM interaction proportional to the sum of dark energy and dark matter densities}

Using (\ref{rw}) the equations (\ref{eq de}), (\ref{eq dm}), (\ref{dw}) for interaction (\ref{Jsum}) can be reduced to the system of 2 ordinary differential equations:
\begin{eqnarray}
  & & \frac{d\rho_{de}}{da}=-\frac{3}{a}(1+c_{a}^{2}+\xi)\rho_{de}-\frac{3}{a}\xi\rho_{dm}- \nonumber \\
  & & -\frac{3}{a}\rho_{de}^{(0)}(w_{0}-c_{a}^{2}), \label{drdesum} \\
  & & \frac{d\rho_{dm}}{da}=\frac{3}{a}\xi\rho_{de}-\frac{3}{a}(1-\xi)\rho_{dm}, \label{drdmsum}
\end{eqnarray}
which has the exact analytical solution:
\begin{widetext}
\begin{eqnarray}
  & & \rho_{de}=3C_{1}\xi a^{-\frac{3}{2}(2+c_{a}^{2})+\frac{3}{2}\sqrt{c_{a}^{2}(c_{a}^{2}+4\xi)}}+
  \frac{3}{2}C_{2}(c_{a}^{2}+2\xi+\sqrt{c_{a}^{2}(c_{a}^{2}+4\xi)})a^{-\frac{3}{2}(2+c_{a}^{2})-
  \frac{3}{2}\sqrt{c_{a}^{2}(c_{a}^{2}+4\xi)}}- \nonumber \\
  & & -\frac{(1-\xi)\rho_{de}^{(0)}(w_{0}-c_{a}^{2})}{1+c_{a}^{2}(1-\xi)}, \label{de sum} \\
  & & \rho_{dm}=-\frac{3}{2}C_{1}(c_{a}^{2}+2\xi+\sqrt{c_{a}^{2}(c_{a}^{2}+4\xi)})a^{-\frac{3}{2}(2+c_{a}^{2})+\frac{3}{2}\sqrt{c_{a}^{2}(c_{a}^{2}+4\xi)}}-
  3C_{2}\xi a^{-\frac{3}{2}(2+c_{a}^{2})-\frac{3}{2}\sqrt{c_{a}^{2}(c_{a}^{2}+4\xi)}}- \nonumber \\
  & & -\frac{\xi\rho_{de}^{(0)}(w_{0}-c_{a}^{2})}{1+c_{a}^{2}(1-\xi)}, \label{dm sum}\\
  & & C_{1}=-\frac{1}{3\sqrt{c_{a}^{2}(c_{a}^{2}+4\xi)}}\left[\rho_{dm}^{(0)}+\xi\rho_{de}^{(0)}\frac{2(1+w_{0}-\xi c_{a}^{2})+(w_{0}-c_{a}^{2})(c_{a}^{2}+\sqrt{c_{a}^{2}(c_{a}^{2}+4\xi)})}{(1+c_{a}^{2}(1-\xi))(c_{a}^{2}+2\xi+\sqrt{c_{a}^{2}(c_{a}^{2}+4\xi)})}\right], \nonumber \\
  & & C_{2}=\frac{1}{3\sqrt{c_{a}^{2}(c_{a}^{2}+4\xi)}}\left[\rho_{de}^{(0)}\frac{1+w_0(1-\xi)}{1+c_{a}^{2}(1-\xi)}
  +2\xi\frac{\rho_{dm}^{(0)}+\frac{\xi\rho_{de}^{(0)}(w_{0}-c_{a}^{2})}{1+c_{a}^{2}(1-\xi)}}{c_{a}^{2}+2\xi+\sqrt{c_{a}^{2}(c_{a}^{2}+4\xi)}}\right]. \nonumber
\end{eqnarray}
\end{widetext}
\begin{figure*}
\centering
\includegraphics[width=0.41\textwidth]{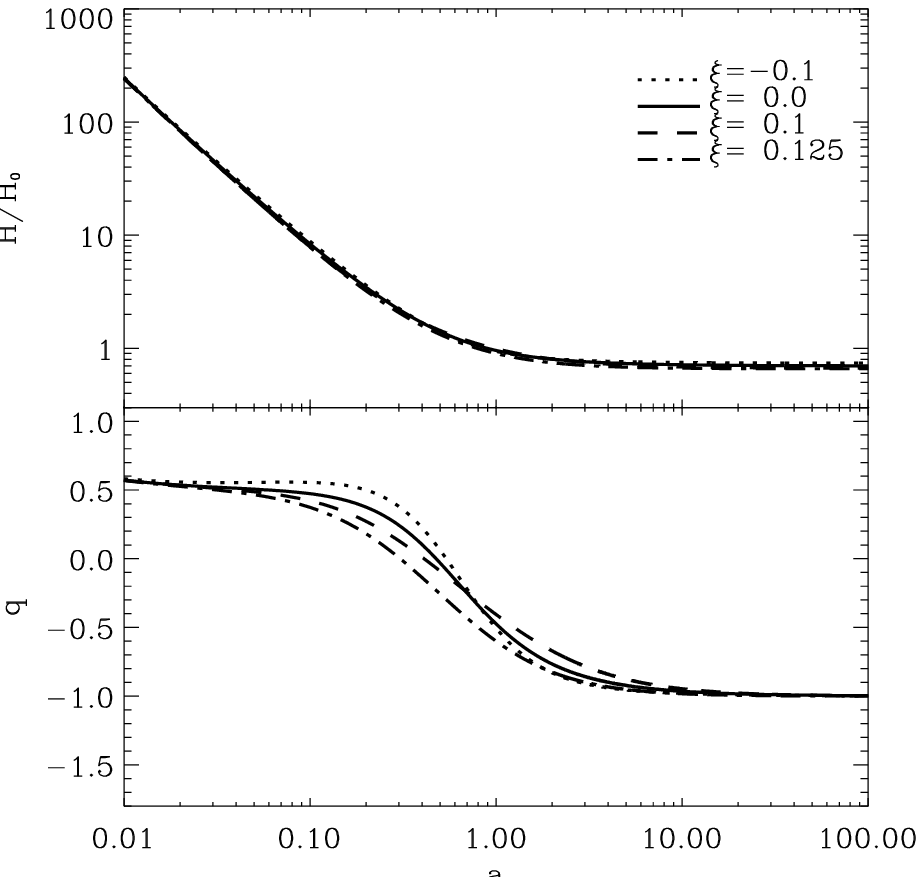}
\includegraphics[width=0.41\textwidth]{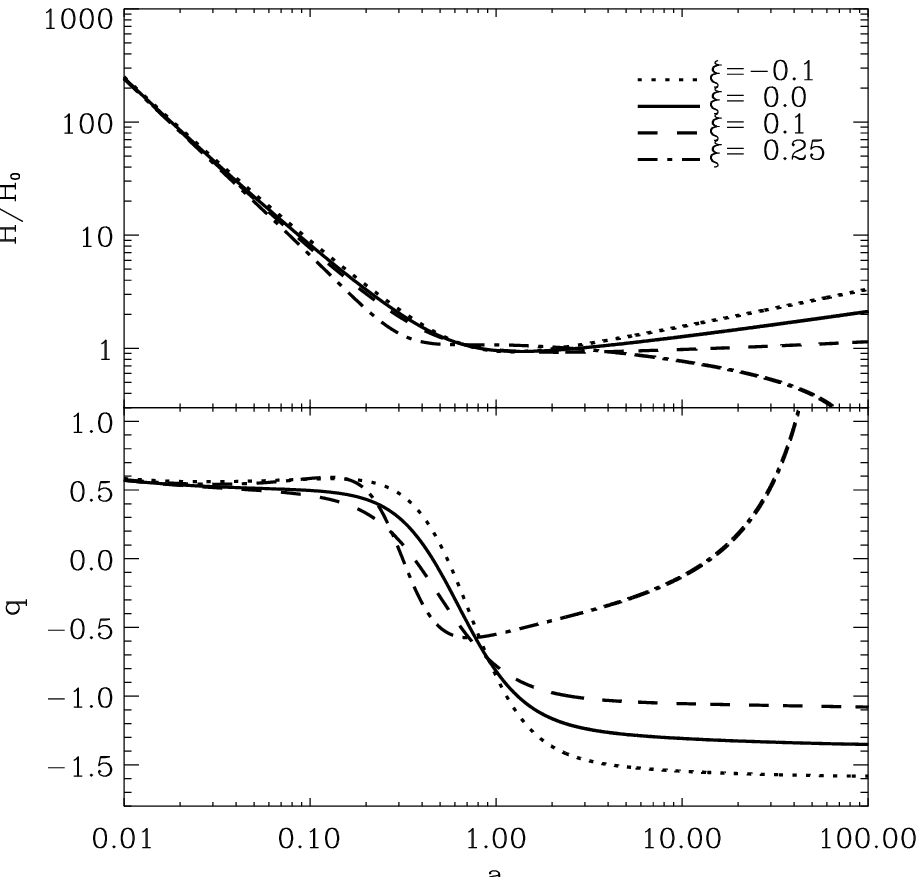}
\caption{Influence of the coupling parameter $\xi$ on dependences of the Hubble parameter $H$ and the deceleration parameter $q$ on $a$. Left -- model with quintessence, right -- with phantom. Cosmological parameters are the same as in Fig. \ref{fig:xm sum1}.}
\label{fig:xm sum2}
\end{figure*}
From (\ref{de sum}), (\ref{dm sum}) we see that in absence of the interaction ($\xi =0$) these expressions turn into the well-known ones for the densities of dark components in the minimally coupled case \cite{Novosyadlyj2010}:
\begin{eqnarray}
 &&\rho_{de}^{(mc)}(a)=\rho_{de}^{(0)}\frac{(1+w_{0})a^{-3(1+c_{a}^{2})}-w_{0}+c_{a}^{2}}{1+c_{a}^{2}}, \label{rode0}
\end{eqnarray}
\begin{eqnarray}
 &&\rho_{dm}=\rho_{dm}^{(0)}a^{-3}.\label{rodm0}
\end{eqnarray}

The density of dark matter during the expantion of the Universe is always positive decreasing function. The density of dark energy is always positive function when $c_a^2>w_0$ for quintessence (for $w_0>-1$ it is the monotonically decreasing function of $a$) and $c_a^2<w_0$ for phantom (for $w_0<-1$ it is the monotonically increasing function of $a$).

The introduction of additional interaction expands substantially the choice of possibilities for evolution of both components. For $\xi\ne0$ and certain values of the parameters $w_{0},\,c_{a}^{2},\,\rho_{de}^{(0)},\,\rho_{dm}^{(0)}$ the energy density of both dark energy and dark matter may become negative during a certain period of evolution of the Universe or oscillate with the change of sign. We consider this behavior of densities to be unphysical and find the range of values of the interaction parameter for which values of the densities of dark matter and dark energy are always positive monotone functions of the scale factor $a$.

First of all, it is known that the set of observational data on the cosmic microwave background anisotropy and the large scale structure of the Universe prefers the close to (\ref{rodm0}) law of the dark matter density change in the past (e.g., \cite{Novosyadlyj2010, Novosyadlyj2011, Novosyadlyj2012, Novosyadlyj2013, Novosyadlyj2014, Sergijenko2011, Sergijenko2015}). From the analysis of expressions in the exponents of $a$ in (\ref{dm sum}) first condition for the coupling parameter follows:
\begin{equation}
 |\xi|\ll\frac{1}{2}. \label{uxi1}
\end{equation}
The range of values of the parameter $\xi$, for which $\rho_{dm}(a)\ge0$ and $\rho_{de}(a)\ge0$ for $0\le a\le\infty$, can be found either from the analysis of expressions
(\ref{de sum})-(\ref{dm sum}) or from the analysis of stability of equations (\ref{drdesum}) and (\ref{drdmsum}) by the method of critical points \cite{Copeland1998} in the phase space $(\rho_{de}, \rho_{dm})$. Both approaches constrain the range of values of $\xi$ to positive ones:
\begin{equation}
 \xi\ge0. \label{uxi2}
\end{equation}
This condition for coupling (\ref{Jsum}) was expected due to the analysis \cite{Neomenko2016} for this kind of interaction.

The condition for absence of oscillations is obtained from the requirement of inequality of the expressions under radicals in the exponents of $a$ in (\ref{de sum})-(\ref{dm sum}):
\begin{equation}\label{ccr}
  \xi\le-\frac{c_{a}^{2}}{4}.
\end{equation}

\begin{figure*}
\centering
\includegraphics[width=0.41\textwidth]{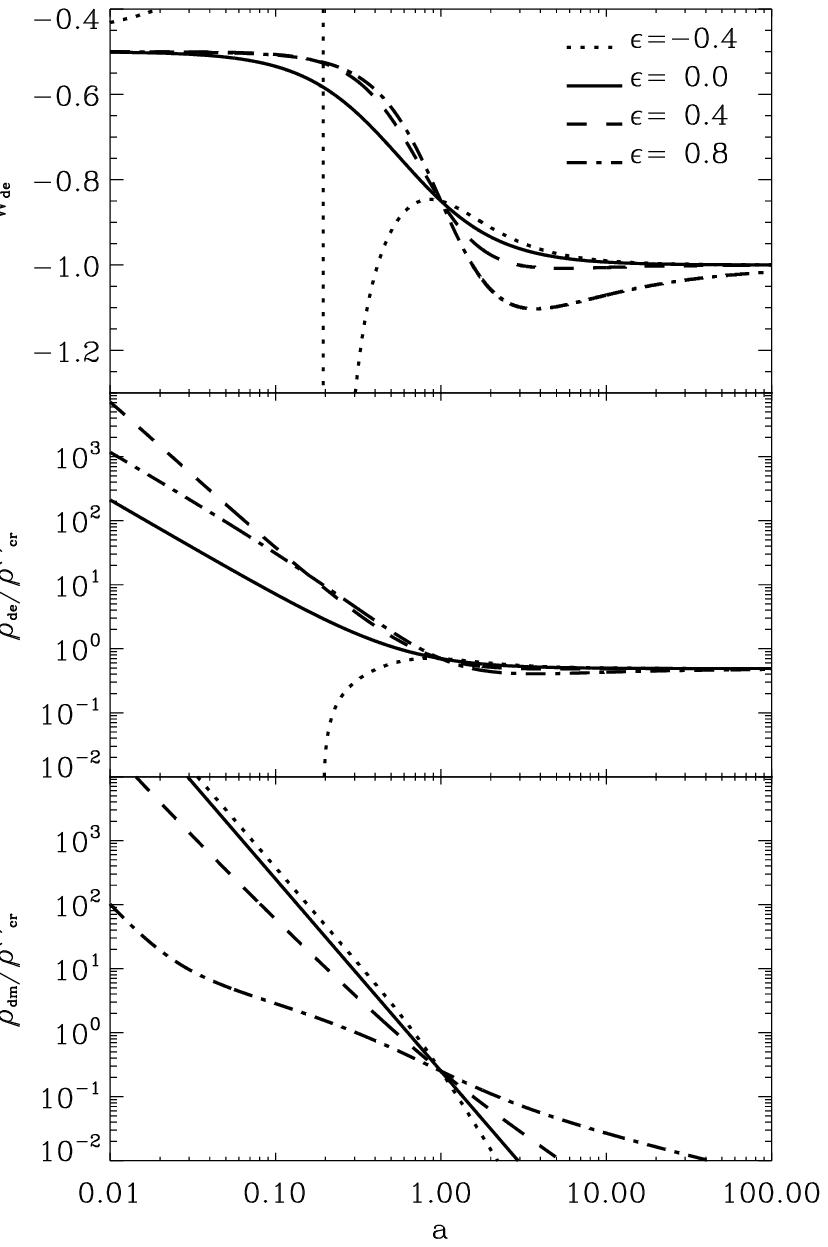}
\includegraphics[width=0.41\textwidth]{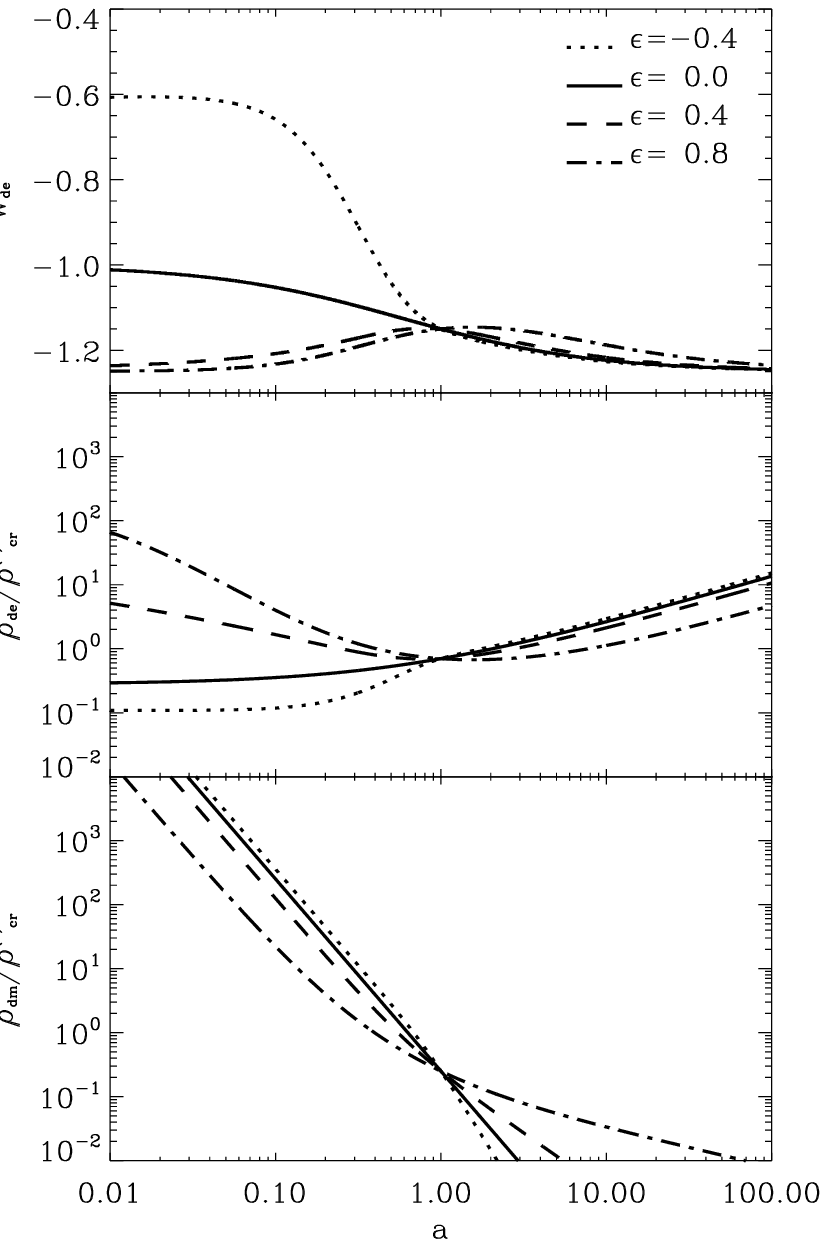}
\caption{Influence of the coupling parameter $\epsilon$ on dependences of the EoS parameter $w$ and the dark energy and dark matter densities $\rho_{de}$, $\rho_{dm}$ on $a$. Left --model with quintessence ($w_{0}=-0.85$, $c_{a}^{2}=-0.5$), right -- with phantom ($w_{0}=-1.15$, $c_{a}^{2}=-1.25$). Cosmological parameters are the same as in Fig. \ref{fig:xm sum1}.}
  \label{fig:xm pro1}
\end{figure*}

We see that for $c_{a}^{2}<0$, $c_{a}^{2}>-{1}/{(1-\xi)}$ the dark energy density is the monotonically decreasing function $a$ with asymptotes $\rho_{de}\rightarrow\infty$ when $a\rightarrow 0$ and $\rho_{de}\rightarrow {(\xi-1)\rho_{de}^{(0)}(w_{0}-c_{a}^{2})}/{[1+c_{a}^{2}(1-\xi)]}$ when $a\rightarrow\infty$. The dark matter density in the case of $\xi\ne0$ has such asymptotes when $a\rightarrow\infty$: $\rho_{dm}\rightarrow -{\xi\rho_{de}^{(0)}(w_{0}-c_{a}^{2})}/{[1+c_{a}^{2}(1-\xi)]}$. The dark energy density has decreased by this value. That is, the coupling (\ref{Jsum}) results in pumping of the energy density from dark energy to dark matter. When $a\rightarrow0$ the asymptote of $\rho_{dm}$ is close to (\ref{rodm0}).

If $c_{a}^{2}<0$, $c_{a}^{2}<-{1}/{(1-\xi)}$ then the density of dark energy is a function of $a$ which varies nonmonotonically with asymptotes $\rho_{de}\rightarrow\infty$ when $a\rightarrow\infty$ and $\rho_{de}\rightarrow\infty$ when $a\rightarrow0$. The density of dark matter in model of the Universe with such dark energy varies according to the same law. In the current and past epochs the dependence of densities of the dark components on $a$ is decreasing and close to (\ref{rodm0}). However, in the future decreasing of the dark energy and dark matter density will stop and they will start to grow with growth of the expansion rate of the Universe. Thus, in the case of this dark energy model with additional coupling to dark matter (\ref{Jsum}) the Big Rip singularity is approached along with a catastrophic increase in the density of dark matter.

In Fig. \ref{fig:xm sum1} the dependences of the EoS parameter $w$ and the densities of dark energy and dark matter $\rho_{de}$, $\rho_{dm}$ calculated on basis of (\ref{rw}) and (\ref{de sum})-(\ref{dm sum}) for cosmological models with $\Omega_{de}=0.7$, $\Omega_{dm}=0.25$, $\Omega_b=0.05$, $\Omega_r=4.17\cdot10^{-5}/h^2$ ($h\equiv H_0/100$ km$\cdot$с$^{-1}\cdot$Mpc$^{-1}$=0.7) and different values of the coupling parameter $\xi$ for quintessence ($w_{0}=-0.85$, $c_{a}^{2}=-0.5$) and phantom ($w_{0}=-1.15$, $c_{a}^{2}=-1.25$) are shown. The case with $\xi <0$ illustrates the change of signs of the dark energy and dark matter densities. In the case of phantom dark energy it is also shown that for large values of the coupling parameter, with violation of the condition (\ref{uxi1}), the solutions (\ref{de sum})-(\ref{dm sum}) give negative values of the dark energy and dark matter densities. Presented dependences show that the evolution of densities $\rho_{de}$, $\rho_{dm}$ in the past is sensitive to the value of coupling parameter $\xi$.

In Fig. \ref{fig:xm sum2} the dependences of the Hubble parameter $H$ and the deceleration parameter $q$ are shown for the cosmological models with quintessence and phantom with the same parameters as in Fig. \ref{fig:xm sum1}.
We see that in the model with quintessence with the coupling parameters $\xi =-0.1, \ 0, \ 0.1, \ 0.125$ $H\rightarrow\infty$, $q\rightarrow 1$ when $a\rightarrow 0$ and $H\rightarrow\left[-{\rho_{de}^{(0)}(w_{0}-c_{a}^{2})}/{(1+c_{a}^{2}(1-\xi))}\right]^{1/2}$, $q\rightarrow -1$ when $a\rightarrow\infty$. For phantom dark energy with the coupling parameters $\xi =-0.1, \ 0.1$ $H\rightarrow\infty$, $q\rightarrow 1$ when $a\rightarrow 0$ and $H\rightarrow\infty$, $q\rightarrow 1/2+3c_a^2/2+3\xi c_a^2/[c_{a}^{2}+\sqrt{c_{a}^{2}(c_{a}^{2}+4\xi)}]$ when $a\rightarrow\infty$. For the model without interaction $\xi =0$ the asymptotes are the same, but $q\rightarrow {1}/{2}+{3c_{a}^{2}}/{2}$ when $a\rightarrow\infty$. For the model with coupling parameter $\xi=0.25$ $H\rightarrow -\infty$, $q\rightarrow +\infty$ when $a\rightarrow\infty $.
So we see that there is a fundamental opportunity to detect an additional interaction between dark components using precision measurements of the Hubble and deceleration parameters.

\section{DE-DM interaction proportional to the product of dark energy and dark matter densities}
\begin{figure*}
\centering
\includegraphics[width=0.41\textwidth]{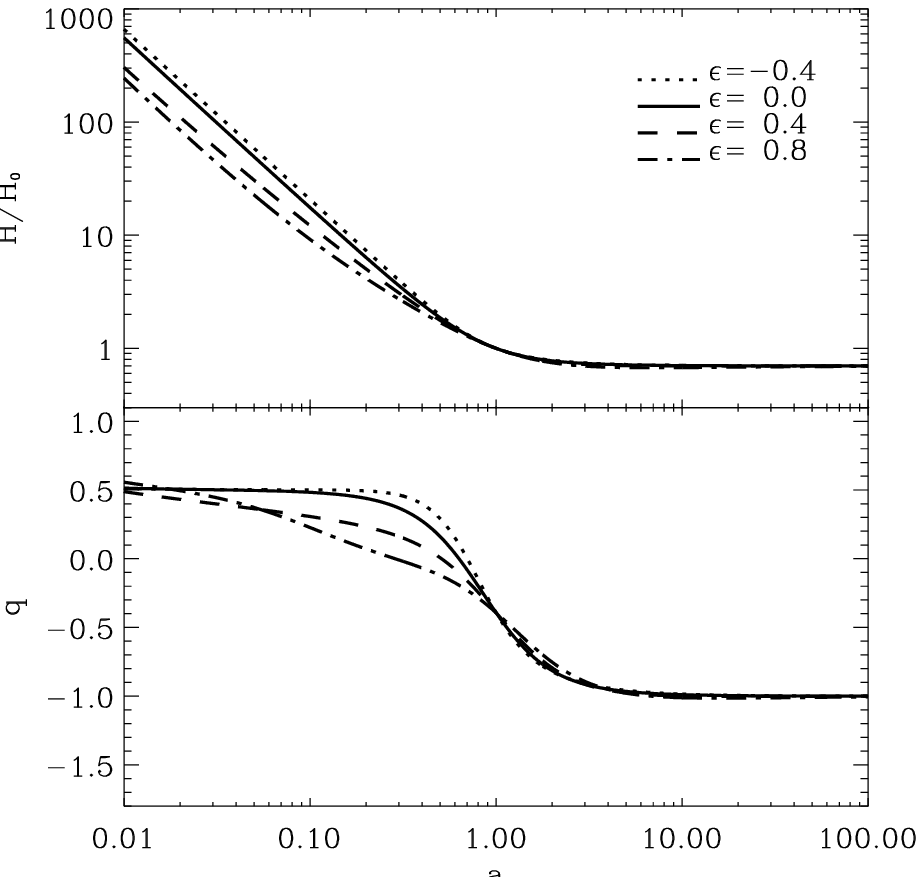}
\includegraphics[width=0.41\textwidth]{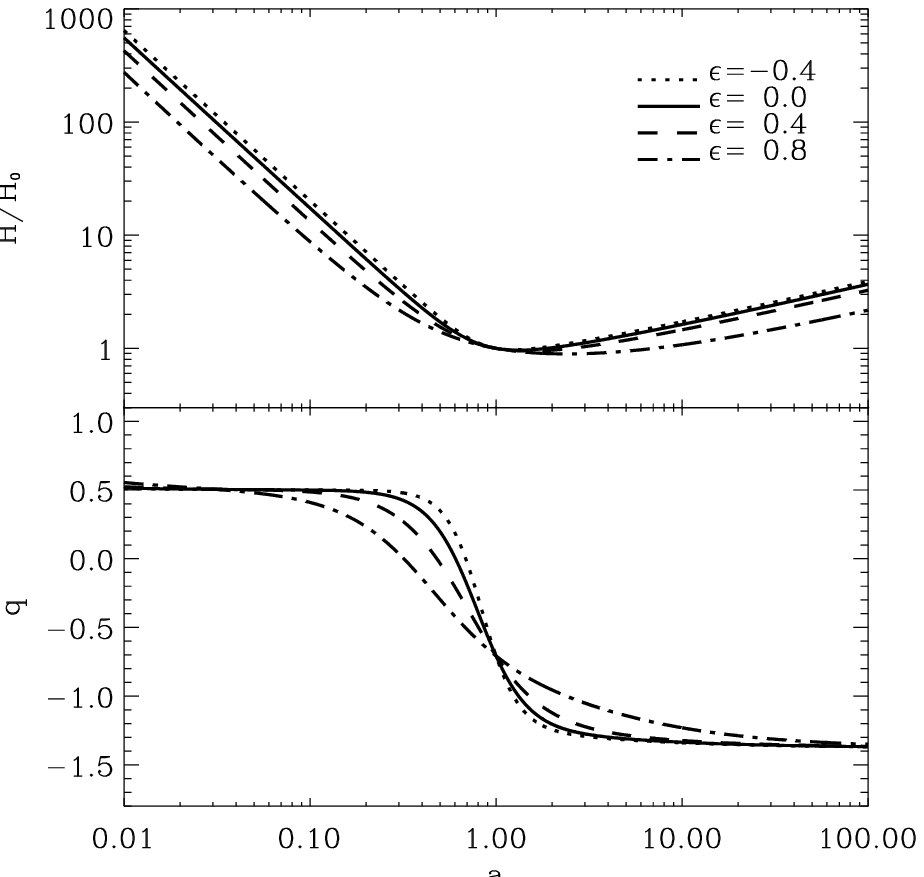}
\caption{Influence of the coupling parameter $\epsilon$ on dependences of the Hubble parameter $H$ and the deceleration parameter $q$ on $a$. Left -- model with quintessence, right -- with phantom. Cosmological parameters are the same as in Fig. \ref{fig:xm sum1}.}
\label{fig:xm prohq1}
\end{figure*}

Let us now consider the dynamics of expansion of the Universe and the evolution of densities of dark energy and dark matter with the interaction proportional to the product of their energy densities (\ref{Jpro}). In this case the equations (\ref{eq de}), (\ref{eq dm}), (\ref{dw}) using (\ref{rw}) are reduced to the system of 1st order ordinary differential equations:
\begin{eqnarray}
  & & \frac{d\rho_{de}}{da}=-\frac{3}{a}(1+c_{a}^{2})\rho_{de}-\frac{3\epsilon}{a\tilde{\rho}}\rho_{de}\rho_{dm}- \nonumber
\end{eqnarray}
\begin{eqnarray}
  & & -\frac{3}{a}\rho_{de}^{(0)}(w_{0}-c_{a}^{2}), \label{drdepro}\\
  & & \frac{d\rho_{dm}}{da}=-\frac{3}{a}\rho_{dm}+\frac{3\epsilon}{a\tilde{\rho}}\rho_{de}\rho_{dm}. \label{drdmpro}
\end{eqnarray}
The product of densities of the dark components in this interaction is divided by $\tilde{\rho}$ so that the interaction (\ref{Jpro}) has the correct dimension.

We can choose $\tilde{\rho}$ as follows:
\begin{eqnarray}
  & & \tilde{\rho}=\rho_{cr}=3H_0^2/8\pi G, \label{cr} \\
  & & \tilde{\rho}=3H^2/8\pi G, \label{tot} \\
  & & \tilde{\rho}=\rho_{de}+\rho_{dm}. \label{sum}
\end{eqnarray}

It should be noted that for the model (\ref{cr}), even for very small $\epsilon$, due to the quadraticity with respect to density, the interaction grows faster than the growth of densities of the dark components. Therefore, for $\rho_{de}, \ \rho_{dm}$ comparable with $\tilde{\rho}/\epsilon$ the interaction is very large leading to the strong increase in the rate of pumping of the energy density from one dark component to the other, hence, the strong deviations in the evolution of densities of the dark components and the dynamics of expansion of the Universe from the model without interaction. Only for $|\epsilon|\ll 1$, when $a$ is approaching $1$, the evolution of densities of the dark components becomes close to the non-interacting case.

Since in the matter and dark energy dominated epocs the contribution of densities of the baryonic and relativistic components in the denominator $\tilde{\rho}$ of model (\ref{tot}) is small and in the radiation dominated epoch the main contribution to dynamics of expansion comes from the energy density of relativistic component, the expansion dynamics in models of the nonlinear interaction (\ref{tot}) and (\ref{sum}) is similar during the whole evolution of the Universe.

Henceforth we consider only the interaction (\ref{tot}). Due to the lack of exact analytical solutions, we integrate the system of equations (\ref{drdepro})-(\ref{drdmpro}) numerically using the Runge-Kutta method implemented in the code dverk.f \cite{dverk} which is publicly available. The results are presented in Fig. \ref{fig:xm pro1} showing the dependences of $w$, $\rho_{de}$ and $\rho_{dm}$ on $a$ for the same values of cosmological parameters as in the previous section.

For the interaction with (\ref{tot}) the rapid growth of product of $\rho_{de}$ and $\rho_{dm}$ with the growth of these densities is compensated by the growth of $\tilde{\rho}$. For small $\epsilon$ the behavior of densities of the dark components is slightly different from the non-interacting model. Only for a very large $\epsilon$ the nonlinearity of this interaction manifests itself. In Fig. \ref{fig:xm pro1} we see that for $\epsilon=-0.4, \ 0, \ 0.4, \ 0.8$ the dark energy with parameters $w_{0}=-0.85$, $c_{a}^{2}=-0.5$ is quintessence and with parameters $w_{0}=-1.15$, $c_{a}^{2}=-1.25$ is phantom. When $\epsilon=-0.4$ the density of quintessence in the early epoch is negative.

As in the case of minimally coupled dark energy ($\epsilon=0$) \cite{Novosyadlyj2010,Novosyadlyj2012}, when $\epsilon=-0.1, \ 0.01, \ 0.1$ the density of dark energy during expansion of the Universe becomes negative if $w_{0}>c_{a}^{2}$ for quintessence ($w_{0}=-0.85$, $-1<c_{a}^{2}<-1/3$) and $w_{0}<c_{a}^{2}$ for phantom ($w_{0}=-1.25$, $-2<c_{a}^{2}<-1$).

For the same cosmological parameters in Fig. \ref{fig:xm prohq1} it is shown how the interaction (\ref{Jpro}) affects evolution of the Hubble parameter $H$ and the deceleration parameter $q$. For this model with small $|\epsilon|$ the influence of interaction on the dynamics of expansion of the Universe is small.

In general, for the quintessence with $w_{0}=-0.85$, $c_{a}^{2}=-0.5$ and $\epsilon\leq-0.212$ the density of dark energy in the past at $a\geq10^{-5}$ was negative and the transition to positive values occurs the later, the larger $|\epsilon|$ is: when $\epsilon=-0.212$, $\rho_{de}$ becomes positive at $a\simeq0.001$, and when $\epsilon=-1$, it becomes positive only at $a=0.439$. For larger values of the coupling parameter ($-0.212<\epsilon\leq1$) the densities of dark energy and dark matter are positive for all $a$ from $10^{-5}$ to $10^{5}$. For the phantom dark energy with $w_{0}=-1.15$, $c_{a}^{2}=-1.25$ $\rho_{de}$ and $\rho_{dm}$ are positive for all $\epsilon$ from -1 to 1 and $a$ from $10^{-5}$ to $10^5$. The total density of the Universe is positive for all values of the coupling parameter from -1 to 1 and the scale factor from $10^{-5}$ to $10^5$ for both quintessence and phantom.

\section*{Conclusions}
In this paper we have analyzed the model of dynamical nonminimally coupled dark energy with EoS parameterized by the value of adiabatic sound speed in the homogeneous and isotropic Universe. Two types of the interaction between dark energy and dark matter depending on the densities of both dark components are considered: proportional to the sum of densities of the dark components and to their product. In both cases for certain parameters the energy densities of dark components may become negative. For the first type of interaction it has been determined that this occurs for a negative value of the coupling parameter. It is also shown that the coupling parameter should be $\ll0.5$ for consistency with the observational data. For the second type of interaction it has been found that for quintessence with the coupling parameter $\lesssim-0.2$ the density of dark energy was negative at the early stages of evolution of the Universe, while for phantom it remains positive throughout the evolution of the Universe for the values of coupling parameter from -1 to 1. In general, in this case for the small absolute values of coupling parameter the dynamics of expansion of the Universe in the past is close to the expansion dynamics in the model with minimally coupled dark energy.

Thus, for both considered types of the interaction between dark energy and dark matter (\ref{Jsum}), (\ref{Jpro}) there is a range of ​​values ​​of the dimensionless coupling parameter, for which the dynamics of expansion of the Universe does not contradict the observational data. Whether the value of this parameter is zero -- no interaction -- or not -- there is an additional interaction -- can be determined by comparing the predictions of different models with the corresponding observational data that are sensitive to such interaction. The search for tests for distinguishing between the minimally coupled models of dark energy and models with the additional interaction is an important task of modern cosmology.

\section*{Acknowledgments}
This work was supported by the project of Ministry of Education and Science of Ukraine (state registration number 0116U001544).

\end{document}